\documentclass[preprint, showpacs, aps, draft]{revtex4}
\usepackage{bm}

\begin{document}
\title{Quantum theory from the geometry of evolving probabilities\let\thefootnote\relax\footnotetext{Presented at MaxEnt 2011, the 31st International Workshop on Bayesian Inference and Maximum Entropy Methods in Science and Engineering, July 10-15, 2011, Waterloo, Canada.}}

\author{Marcel Reginatto}
\affiliation{Physikalisch-Technische Bundesanstalt, Bundesallee 100,
38116 Braunschweig, Germany}

\author{Michael J. W. Hall}
\affiliation{Theoretical Physics, RSPE, Australian National University, Canberra ACT 0200, Australia}
\affiliation{Centre for Quantum Dynamics, Griffith University, Brisbane, QLD 4111, Australia\\}

\begin{abstract}
We consider the space of probabilities $\{P(x)\}$, where the $x$ are coordinates of a configuration space. Under the action of the translation group, $P(x) \rightarrow P(x+\theta)$, there is a natural metric over the parameters $\theta$ given by the Fisher-Rao metric. This metric induces a metric over the space of probabilities. Our next step is to set the probabilities in motion. To do this, we introduce a canonically conjugate field $S$ and a symplectic structure; this gives us Hamiltonian equations of motion. We show that it is possible to extend the metric structure to the \emph{full space} of the $(P,S)$, and this leads in a natural way to introducing a K\"{a}hler structure; i.e., a geometry that includes compatible symplectic, metric and complex structures.

The simplest geometry that describes these spaces of evolving probabilities has remarkable properties: the natural, canonical variables are precisely the wave functions of quantum mechanics; the Hamiltonian for the quantum free particle can be derived from a representation of the Galilean group using purely geometrical arguments; and it is straightforward to associate with this geometry a Hilbert space which turns out to be the Hilbert space of quantum mechanics. We are led in this way to a reconstruction of quantum theory based solely on the geometry of probabilities in motion.
\end{abstract}


\maketitle

\section{Probabilities, translations, and information geometry}\label{ptig}

Our starting point is an $n$-dimensional configuration space, with
coordinates $x\equiv \{x^1,\dots,x^n\}$ and probability densities $P(x)$ satisfying
$P(x) \geq 0$ and $\int d^n x P(x)=1$.

If we consider the action of the translation group $T$ on $P(x)$,
$T:P(x) \rightarrow P(x + \theta)$, there is a natural metric $\gamma_{jk}$ on the space of
parameters $\theta$: the Fisher-Rao metric \cite{Rao87},
\begin{equation}\label{metricTheta}
\gamma_{jk} = \frac{\alpha}{2} \int d^{n}x \frac{1}{P(x + \theta)}
\frac{
\partial P(x + \theta)}{\partial \theta ^{j}}\frac{\partial
P(x + \theta)}{\partial \theta ^{k}},
\end{equation}
where $\alpha$ is a constant. The line element $d{\sigma}^2=\gamma_{jk}\Delta^j \Delta^k$ (where $|\Delta^k|<<1$) defines a distance between the two probability distributions $P(x + \theta)$ and $P(x + \theta + \Delta)$.

It will be convenient to consider another form of the metric.
In particular, using the equality $\frac{\partial P(x+\theta)}{\partial
\theta^j}=\frac{\partial P(x+\theta)}{\partial x^j}$, and making the change of integration variables $x \rightarrow x + \theta$, the metric is proportional to the Fisher information matrix,
\begin{equation}\label{metricX}
\gamma_{jk} = \frac{\alpha}{2} \int d^{n}x \frac{1}{P(x)}
\frac{\partial P(x)}{\partial x^{j}} \frac{\partial
P(x)}{\partial x^{k}}.
\end{equation}

It follows that the line element $d{\sigma}^2=\gamma_{jk}\Delta^j \Delta^k$ induces a
line element $ds^2$ in the space of probability densities. Introducing the notation $P_x=P(x)$, $\delta P_x\equiv ~ \frac{\partial P(x)}{\partial x^{j}}\Delta^j$, we have
\begin{equation}\label{Jmetric}
ds^2= \frac{\alpha}{2} \int d^{n}x \frac{1}{P_x}\delta P_x\, \delta P_x =\int d^nx\,d^nx' \,g_{PP}(x,x')\, \delta P_x\,\delta P_{x'}~,
\end{equation}
The equivalence of the two line elements can be checked by direct substitution. Eq. (\ref{Jmetric}) was introduced by Jeffreys \cite{J46}. The induced metric $g_{PP}$ is diagonal, and given by
\begin{equation}\label{metricP}
g_{PP}(x,x')=\frac{\alpha}{2 P_x} \delta(x-x').
\end{equation}

\section{Symplectic geometry and observables}

We now set the probabilities in motion. To do this, we assume the dynamics of $P(x)$ are generated by an action principle.  Hence, we introduce an
auxiliary field $S$ which is canonically conjugate to $P$, and a
corresponding Poisson bracket for any two functionals $F[P,S]$ and $G[P,S]$:
\begin{equation}\label{PoissonBrackets}
\left\{ F,G\right\} =\int d^{n}x \left\{
\frac{\delta F}{\delta P} \frac{\delta G}{\delta S}
 - \frac{\delta F}{\delta S}  \frac{\delta G}{\delta P} \right\}.
\end{equation}
The equations of motion for $P$ and $S$ then have the form
\begin{equation} \label{eqmotion}
\dot{P} = \left\{ P,{\cal H}\right\} =\frac{\delta
{\cal H}}{\delta S},~~~~~
\dot{S} = \left\{ S,{\cal H}\right\} =-\frac{\delta
{\cal H}}{\delta P},
\end{equation}\nonumber
where ${\cal H}$ is the ensemble Hamiltonian that generates time translations.

As is well known, the Poisson bracket can be rewritten geometrically as
\begin{equation}
\left\{ F,G\right\} =\int d^{n}x \,d^nx'\, \left( \delta F/\delta P_x\, , \; \delta F/\delta S_x\right) \, \Omega(x,x') \, \left(
\begin{array}{c}
\delta G/\delta P_{x'} \\
\delta G/\delta S_{x'}
\end{array}
\right) ,
\end{equation}\nonumber
where $\Omega$ is the corresponding symplectic form, given in this case by
\begin{equation} \label{Omega}
\Omega(x,x')=\left(
\begin{array}{cc}
0 & 1 \\
-1 & 0
\end{array}
\right)\delta(x-x')~.
\end{equation}
We thus have a symplectic structure and a corresponding {\it symplectic geometry\/} \cite{A89}.

The fundamental variables of our phase space are the probabilities $P_x$ and the auxiliary function $S_x$. We now introduce the notion of an {\it observable\/} on this phase space, as any functional $A[P,S]$ that satisfies certain requirements. For example, the infinitesimal
canonical transformation generated by any observable $A$ must
preserve the normalization and positivity of $P$. This implies the
two conditions \cite{HR05}
\begin{equation}
A[P,S+c] = A[P,S],~~~~~\delta A / \delta S = 0 ~ \textrm{if} ~
P(x)=0.
\end{equation}
Note that the first condition implies gauge invariance of the theory under $S\rightarrow S + c$. A more general condition that might be imposed on observables, which leads to a natural statistical interpretation, is that they are homogeneous of degree one with respect to $P$,
\begin{equation}\label{homogeneityOne}
A[\lambda P, S] = \lambda A[P,S].
\end{equation}
If we now differentiate both
sides of Eq. (\ref{homogeneityOne}) with respect to $\lambda$ and
set $\lambda=1$, we get \cite{H08}
\begin{equation}\label{homogeneity}
A[P, S] = \int d^n x \, P(\delta A / \delta P) := \langle
\delta A / \delta P \rangle,
\end{equation}
i.e., $A$ can be calculated by integrating over a local density.
The main motivation for introducing the homogeneity condition is that it
consistently allows observables to be interpreted both as generators
of canonical transformations and as expectation values.

It is possible to give a physical interpretation to the canonically
conjugate variable $S$. Notice that $\int dx\, P\nabla S$ is the
canonical infinitesimal generator of translations, since
\begin{eqnarray} \label{Ptranslation}
\delta P(x) = \delta \textbf{x} \cdot \left \{ P, \int dx\,
P\nabla S \right \} = - \delta \textbf{x} \cdot \nabla P ,
\\ \label{Stranslation}
\delta S(x) = \delta \textbf{x} \cdot \left \{ S, \int dx\,
P\nabla S \right \} = - \delta \textbf{x} \cdot \nabla S ,
\\\nonumber
\end{eqnarray}
under action of the generator, and therefore $P\nabla S$ can be
considered a {\it local momentum density\/} \cite{RH09}.

\section{K\"{a}hler geometry}

We now want to consider the following question: Can we extend the metric $g_{PP}(x,x')$ in Eq.~(\ref{metricP}), which is only defined on the subspace of probabilities $P$, to the full phase space of $P$ and $S$? It can be done, but certain conditions which ensure the compatibility of the metric and symplectic structures have to be satisfied (see the Appendix for a proof). These conditions amount to requiring that the space have a K\"{a}hler structure. We show in this way the beautiful result that the natural geometry of the space of probabilities in motion is a {\it K\"{a}hler geometry\/}.

A K\"{a}hler structure brings together metric, symplectic and complex structures in a harmonious way. To define such a space, introduce a complex structure $J_{\ b}^{a}$ and impose the following conditions \cite{G82},
\begin{eqnarray}
\Omega _{ab} &=& g_{ac}J_{\ b}^{c} \;,  \label{c1}\\
J_{\ c}^{a}g_{ab}J_{\ d}^{b} &=& g_{cd} \;,  \label{c2}\\
J_{\ b}^{a}J_{\ c}^{b} &=& -\delta _{\ c}^{a} \;.  \label{c3}
\end{eqnarray}
Eq. (\ref{c1}) is a compatibility equation between $\Omega _{ab}$ and $g_{ab}$, Eq. (\ref{c2}) is the condition that the metric should be Hermitian, and Eq. (\ref{c3}) is the condition that $J_{b}^{a}$ should be a complex structure. We will now derive the local solutions to these equations.

We saw before that the metric
over the subspace of probabilities is diagonal and given by
$g_{PP}(x,x')=\frac{\alpha}{2P_x} \delta(x-x')$. We assume that
the full metric $g_{ab}$ is also diagonal; that is, of the form
$g_{ab}(x,x')=g_{ab}(x)\delta(x-x')$ (this assumption corresponds to a locality assumption). Then $g_{ab}$ is a
real, symmetric matrix of the form
\begin{equation}
g_{ab}=\left(
\begin{array}{cc}
\frac{\alpha}{2 P_x} & \quad g_{PS} \quad \\
g_{SP} & \quad g_{SS} \quad
\end{array}
\right)\delta(x-x') .
\end{equation}
The elements $g_{PS}=g_{SP}$ and $g_{SS}$ still need to be
determined.

Since $\Omega _{ab}(x,x')$ in Eq.~(\ref{Omega}) is also diagonal, Eq. (\ref{c1}) implies that $J_{\
b}^{c}(x,x')$ is diagonal; i.e. $J_{\ b}^{c}(x,x')=J_{\
b}^{c}(x)\delta(x-x')$. Using Eq.
(\ref{c3}), one can show that $J_{\ b}^{c}$ depends on two arbitrary
functionals (which we write as $A_x$ and $C_x$ in the equation below)
and can be written in the form
\begin{equation}\label{formOfJ}
J_{\ b}^{c}=\left(
\begin{array}{cc}
A_x & C_x (1+A_x^2)  \\
-\frac{1}{C_x} & -A_x
\end{array}
\right)\delta(x-x') .
\end{equation}
It is not difficult to get expressions for $g_{ab}(x,x')$ and $J_{\ b}^{c}(x,x')$ from the two remaining equations, Eqs. (\ref{c1}) and (\ref {c2}). These final expressions depend only on the arbitrary functional $A_x$, with
\begin{equation}\label{g_general}
g_{ab}=\left(
\begin{array}{cc}
\frac{\alpha}{2 P_x} & A_x \\
A_x &  \frac{2 P_x}{\alpha}(1+A_x^{2})
\end{array}
\right)\delta(x-x') ,
\end{equation}
\begin{equation}\label{j_general}
J_{\ b}^{a}=\left(
\begin{array}{cc}
A_x &  \frac{2 P_x}{\alpha}(1+A_x^{2}) \\
-\frac{\alpha}{2 P_x} & -A_x
\end{array}
\right)\delta(x-x') .
\end{equation}

\section{Complex coordinates}

Different choices of $A_x$ in Eqs. (\ref{g_general}) and (\ref{j_general}) correspond in general to different K\"{a}hler geometries. From the mathematical point of view, the simplest one among these is a {\it flat\/} K\"{a}hler space, and it follows from the simplest choice, $A_x=0$. To show this, we carry out a complex transformation. We set $A_x=0$ and consider the K\"{a}hler structure given (up to a product with $\delta(x-x')$) by
\begin{equation}\label{OgJPS}
\Omega _{ab}=\left(
\begin{array}{cc}
0 & 1 \\
-1 & 0
\end{array}
\right),  ~~~~~
g_{ab}=\left(
\begin{array}{cc}
\frac{\alpha}{2P_x} & 0 \\
0 & \frac{2P_x}{\alpha}
\end{array}
\right),  ~~~~~
J_{\ b}^{a}=\left(
\begin{array}{cc}
0 & \frac{2P_x}{\alpha} \\
-\frac{\alpha}{2P_x} & 0
\end{array}
\right).  \label{SPogj}
\end{equation}
The complex coordinate transformation that is required is nothing
but the Madelung transformation, $\psi =\sqrt{P}\exp (iS/\alpha)$, $\psi ^{\ast }=\sqrt{P}\exp (-iS/\alpha)$. In terms of the new variables, Eqs. (\ref{SPogj}) take the standard flat-space form \cite{G82}
\begin{equation}
\Omega _{ab}=\left(
\begin{array}{cc}
0 & i\alpha \\
-i\alpha & 0
\end{array}
\right),~~~~~
g_{ab}=\left(
\begin{array}{cc}
0 &  \alpha \\
\alpha & 0
\end{array}
\right),~~~~~
J_{\ b}^{a}=\left(
\begin{array}{cc}
-i & 0 \\
0 & i
\end{array}
\right).
\end{equation}
This shows that the simplest geometrical formulation of the space of probabilities in motion has a natural set of fundamental variables -- where, identifying the constant $\alpha$ with $\hbar$, these fundamental variables are precisely the {\it wave functions\/} that we encounter in quantum mechanics. This is a remarkable result, because we have not introduced {\it any} assumptions that concern quantum mechanics. This result was derived using only geometrical assumptions.

\section{The free particle}

We now turn our attention to the description of a free particle using the formalism that we have developed. We consider the case where the configuration space is the Euclidean space $R^3$. This space has a natural metric $h_{ab}$ given by $h_{ab}=\delta_{ab}$ in Cartesian coordinates.

To describe a free particle, we look for a realization of the Galilean group in terms of the algebra of Poisson brackets defined by Eq. (\ref{PoissonBrackets}). The Galilean group has 10 generators:\\
\hspace*{2em} $A_i$ : space displacements,\\
\hspace*{2em} $H$ : time displacements,\\
\hspace*{2em} $L_i$ : space rotations,\\
\hspace*{2em} $ G_i$ : Galilean transformations (``boosts''),\\
where $i=1,2,3$. The generator $H$ transforms as a scalar, while
$A_i$, $L_i$, and $G_i$ transform as vectors. These generators
have to satisfy the Poisson bracket relations \cite{Finkelstein73}
\begin{eqnarray}
\{H,A_i\} &=& 0,\label{gg1}\\
\{H,L_i\} &=& 0,\label{gg2}\\
\{H,G_i\} &=& -A_i,\label{gg3}\\
\{L_i,A_j\} &=& \epsilon_{ijk}A_k,\label{gg4}\\
\{L_i,L_j\} &=& \epsilon_{ijk}L_k,\label{gg5}\\
\{L_i,G_j\} &=& \epsilon_{ijk}G_k,\label{gg6}\\
\{A_i,A_j\} &=& 0,\label{gg7}\\
\{A_i,G_j\} &=& -m\delta_{ij},\label{gg8}\\
\{G_i,G_j\} &=& 0\label{gg9},
\end{eqnarray}
where $m$ is the mass of the particle.

We represent the generators by observables. For space displacements and rotations one finds, by considering the corresponding infinitesimal transformations of $P$ and $S$, that
\begin{equation}
A_i = \int d^3x \; P \; \left(\partial_i S\right) ,~~~~~
L_i = \int d^3x \; P \; \left(\epsilon_{ijk} \; x_j \; \partial_k S\right),
\end{equation}
up to additive constants (cf. Eqs.~(\ref{Ptranslation}) and (\ref{Stranslation})).  Further, for the Galilean boost transformations it is natural to choose
the observables
\begin{equation}
G_i = \int d^3x \; P \; \left( m x_i - t \partial_i S \right),
\end{equation}
where $t$ is the time. This follows from the standard definition $G_i = (mQ_i - tA_i)$ and the choice $Q_i = \int d^3x \; P \;  x_i$ for the position observable. One can check that Eqs. (\ref{gg4}-\ref{gg9}) are satisfied when we make these choices. Note that the generators satisfy the homogeneity condition, Eq. (\ref{homogeneity}), and have clear interpretations as expectation values.

The one remaining step is to find an observable that satisfies Eqs. (\ref{gg1}-\ref{gg3}). The first two equations, Eqs. (\ref{gg1}-\ref{gg2}), will be satisfied by any scalar $H$, since a scalar is invariant under translations and rotations.
Using $G_k = (mQ_k - tA_k)$ and $\{H,A_k\}=0$, Eq. (\ref{gg3}) then simplifies to
\begin{equation}
\{H,G_i\} = -m\int d^3x \frac{\delta H}{\delta S}x_i   = -A_i = -\int d^3x \; P \; \partial_i S. \label{gg3a}
\end{equation}
Eq. (\ref{gg3a}) does not have a unique solution, and different choices of $H$ will correspond to different theories of the free particle.

It is remarkable that the \emph{only} scalar that can be constructed using purely geometrical quantities \emph{does} satisfy Eq. (\ref{gg3a}).  Therefore, from the point of view of the geometry of probabilities in motion, it is the natural choice for the generator of time translations. It is straightforward to calculate this scalar. Using an argument similar to the one used in the first section of this paper, we can show that the metric over the fields $P$ and $S$ given in Eq. (\ref{OgJPS}) induces a corresponding metric over the space of parameters,
\begin{equation}
g_{jk} = \frac{2}{\alpha}
\int d^3 x \; P \;  \left( \frac{\partial S}{\partial
x^{j}} \frac{\partial S}{\partial x^{k}} + \frac{\alpha^2}{4 P^2}
\frac{\partial P}{\partial x^{j}} \frac{\partial P}{\partial x^{k}}
\right).
\end{equation}
To get a scalar, we need to contract $g_{jk}$ with a tensor with two upper indices. The only geometrical object that we have available for this is the inverse metric of the Euclidean configuration space, $h^{ab}=\delta^{ab}$. If we contract $g_{jk}$ with $h^{ab}$, multiply by the constant $\frac{\alpha}{4m}$, and set $\alpha = \hbar$, the result is
\begin{equation}
H = \frac{\alpha}{4m} \delta^{jk} g_{jk} =  \frac{1}{2m} \int d^n x \;
 \left[ P \left| \nabla S \right|^2  + \frac{\hbar^2}{4 P}
\left| \nabla P \right|^2 \right] = \frac{\hbar^2}{2m} \int d^n x \;
 \left| \nabla \psi \right|^2.
\end{equation}
This is the average energy (and the ensemble Hamiltonian) of a free particle in \emph{quantum mechanics}. This leads therefore to the quantum theory of a free particle.  In particular, Eq.~(\ref{eqmotion}) is equivalent to the Schr\"{o}dinger equation $i\hbar\frac{\partial \psi}{\partial t} = -\frac{\hbar^2}{2m} \nabla^2 \psi$.

Another choice is given by $H$ in the limit where $\hbar \rightarrow 0$. Then we get the average energy (and the ensemble Hamiltonian) of a free particle in \emph{classical mechanics}, and the classical theory of a free particle \cite{HR05}.

\section{The Hilbert space of quantum mechanics from the geometric approach}

There is a standard construction that associates a complex Hilbert space with any infinite dimensional K\"{a}hler space. Given two complex functions $\phi $ and $\varphi$, define the Dirac product by \cite{K79}
\begin{eqnarray}
\langle \phi |\varphi \rangle &=&\frac{1}{2 }\int \left\{ \left(
\phi (x^{\mu }),\phi ^{\ast }(x^{\mu })\right) \cdot \left[
g+i\Omega \right] \cdot \left(
\begin{array}{c}
\varphi (x^{\mu }) \nonumber\\
\varphi ^{\ast }(x^{\mu })
\end{array}
\right) \right\} d^{n}x \\
&=&\frac{1}{2 }\int \left\{ \left( \phi (x^{\mu }),\phi ^{\ast
}(x^{\mu })\right) \left[ \left(
\begin{array}{cc}
0 & 1 \\
1 & 0
\end{array}
\right) +i\left(
\begin{array}{cc}
0 & i \\
-i & 0
\end{array}
\right) \right] \left(
\begin{array}{c}
\varphi (x^{\mu }) \\
\varphi ^{\ast }(x^{\mu })
\end{array}
\right) \right\} d^{n}x \nonumber\\
&=&\int \phi ^{\ast }(x^{\mu })\varphi (x^{\mu })d^{n}x
\end{eqnarray}
In this way, the Hilbert space structure of quantum mechanics follows from the K\"{a}hler geometry. In particular, the complex structure that is needed for the formulation of quantum mechanics arises in a very natural way -- it is forced upon us by the geometry.

\section{Discussion}

We have shown that quantum theory emerges from the geometry of probabilities in motion. This is a reconstruction of quantum theory that seems rather surprising to us, in that none of the elements that are usually assumed to be characteristic of quantum theory (e.g., a representation in terms of an algebra of operators, uncertainty relations, the assumption of a classical theory that is quantized, etc.) are introduced {\it a priori\/}. Instead, our starting points are the natural metric on the space of probabilities (information geometry) and the description of evolution in terms of a Hamiltonian formalism (symplectic geometry), together with requirements of consistency (K\"{a}hler geometry) and simplicity (the choice of a flat K\"{a}hler space).

There is one important additional assumption that needs further discussion, and that is the choice of the translation group when defining the information metric. Since the information metric relates to probabilities, and these probabilities are associated with measurements, it makes sense to look at the role that the translation group plays in measurement theory, in particular as it relates to the {\it limitations\/} imposed by measurement uncertainty. This can provide some insight on why the translation group is important here, even if our remarks are of necessity qualitative rather than quantitative.

It will be useful to go back to one of the oldest concepts in measurement, the idea of \emph{resolution} as it is understood in optics. Here the translation group already makes an appearance. The image of a star, when observed through a telescope with finite resolution, will be blurred. To give an operational definition of the resolution of the telescope, you take two (perhaps hypothetical) point-sources and consider the pair of blurred images. If it is possible to tell them apart, you say that the telescope can resolve them. A measure of resolution is given for example by the well know Rayleigh criterion. It is clear that resolution means the ability to \emph{distinguish} between two point-sources that are close together. There is another operational definition of resolution which is equivalent: take a single point source, consider its image, and then consider the action of the \emph{translation group} on this one image. Define the resolution of your telescope by the smallest distance that you need to move the image such that the superposition of the original image and the displaced image are distinguishable.

In our case, we are dealing with the space of probabilities on an $n$-dimensional configuration space rather than the space of blurred images of a telescope, but there is a clear analogy between both situations. Just as is done in optics, we introduce a concept of resolution in this space by considering the ability to distinguish between a probability $P(x)$ and a probability $P(x+\theta)$ \emph{that results from the action of the translation group}. Since the metric of Eq. (\ref{metricX}) or, equivalently, Eq. (\ref{metricP}), is the one that is associated with the action of the translation group, it will quantify the ability to distinguish between the two in the limit of infinitesimal displacements.

\appendix

\section{Symplectic geometry, compatibility conditions, and K\"{a}hler structure}\label{AppSGCCKS}

We consider a finite space, but similar relations hold for infinite dimensional spaces. A symplectic vector space is a vector space $V$ that is equipped with a bilinear form $\omega : V \times V \rightarrow R$ that is \cite{S90}:\\a. Skew-symmetric: $\omega(u,v) = -\omega(v,u)$ { for all } $u,v ~\epsilon ~V$,\\b. Non-degenerate: if $\omega(u,v) = 0$ { for all } $v~\epsilon ~V,$ { then } $u=0$.\\The standard space is $\Re^{2n}$, and typically $\omega$ is chosen
to be the matrix
\begin{equation}
\omega_{ab} =\left(
\begin{array}{cc}
0 & \textbf{1}^n \\
-\textbf{1}^n & 0
\end{array}
\right),
\end{equation}
where $\textbf{1}^n$ is the unit matrix in $n$ dimensions.

Consider the dual space $V^*$. The symplectic structure can be identified with an element of $V^* \times V^*$, so that $\omega(u,v) = \omega_{ab}u^a v^b$. Since the spaces $V$ and $V^*$ are isomorphic, there is a $\omega^{ab}$ that is the dual of $\omega_{ab}$. This $\omega^{ab}$ can be identified with an element of $V \times V$. The convention is to set \cite{S90}
\begin{equation}
\omega^{ab}
= -(\omega^{-1})^{ab}
=\left(
\begin{array}{cc}
0 & \textbf{1}^n \\
-\textbf{1}^n & 0
\end{array}
\right),
\end{equation}
so that $\omega^{ac}\omega_{cb}=-\delta^{a}_{~b}.$

We assume there is a metric in the space, $g_{ab}=g_{ba}$, and a corresponding inverse metric $g^{ab}$ with $g_{ab}g^{bc}=\delta_{ac}$ (indices are raised and lowered with $g_{ab}$ and $g^{ab}$) The metric also defines a map $V \rightarrow V^*$ to the dual space in an obvious way. Therefore, the space has two linear operators that induce maps $V \rightarrow V^*$, $\omega_{ab}$ and $g_{ab}$. They will be related by an equation of the form $\omega_{ab} = g_{ac} j^c_{~b}$ for some choice of linear operator $j^c_{~b}$. This is Eq. (\ref{c1}), the first of the K\"{a}hler conditions.

The relations $\omega^{ac}\omega_{cb}=-\delta^{a}_{~b}$ and $\omega_{ab} = g_{ac} j^c_{~b}$ lead to the condition
$j^a_{~s} j^s_{~c} = -\delta^{a}_{~c}$, that is, that $j^a_{~b}$ is a complex structure. This is Eq. (\ref{c3}), the third of the K\"{a}hler conditions.

Finally, $\omega_{ab} = g_{ac} j^c_{~b}$ and $j^a_{~s} j^s_{~c} = -\delta^{a}_{~c}$, together with the symmetries $-\omega_{cb} = \omega_{bc}$ and $g_{ba}=g_{ab}$, lead to $g_{cd} = j^a_{~c} g_{ab} j^b_{~d}$ which is Eq. (\ref{c2}), the second of the K\"{a}hler conditions.

This shows that consistency requirements imply that a space with both symplectic and metric structures must have a K\"{a}hler structure.

\end{document}